# A critical point in $Sr_{2-x}IrO_4$ and less distorted $IrO_6$ octahedra induced by deep Sr-vacancies

Jie Cheng[a*], Shengli Liu[a,b*], Bin Li[a*], Haiyun Wang[a], Yu Wang[c] and Wei Xu[d,e]

[a] College of Science, Center of Advanced Functional Ceramics, Nanjing University of Posts and Telecommunications, Nanjing, Jiangsu 210023, China

[b] Nanjing University (Suzhou) High-Tech Institute, Suzhou, 215123, China

[c] Shanghai Synchrotron Radiation Facility, Shanghai Institute of Applied Physics, Chinese Academy of Sciences, Shanghai 201204, China

[d] Beijing Synchrotron Radiation Facility, Institute of High Energy Physics, Chinese Academy of Sciences, Beijing 100049, China

[e] Rome International Center for Materials Science, Superstripes, RICMASS, via dei Sabelli 119A, I-00185 Roma, Italy

E-mail: chengj@njupt.edu.cn, liusl@njupt.edu.cn and libin@njupt.edu.cn

## Abstract

For the $Sr_{2-x}IrO_4$ system, recent work pointed out the abnormal electronic and magnetic properties around a critical point. Here, to understand Sr-vacancy effect on spin-orbit coupled Mott insulator $Sr_2IrO_4$, the electronic structure and local structure distortion for $Sr_{2-x}IrO_4$ system have been investigated by x-ray absorption spectroscopy (XAS). By comparing the intensity of white-line features at the Ir $L_{2,3}$-edge x-ray absorption near-edge spectroscopy (XANES), we observe a sudden rise of the branching ratio in the vicinity of the critical point. Further analysis on the Ir $L_3$-edge extended x-ray absorption fine structure (EXAFS) and calculated data demonstrated that the abrupt enhancement of the branching ratio is intimately related to the less distorted $IrO_6$ octahedra induced by deep Sr-vacancies, which in turn alters the relatively strength of the spin-orbit coupling (SOC) and crystal electric field (CEF) that would dictate the abnormal behavior of electronic and magnetic structure near the critical point.

## 1. Introduction

The 5$d$ transition metal oxides (TMOs) have been the topic of intense research in recent years due to their numerous intriguing phenomena [1, 2]. Because of the large atomic mass associated with 5$d$ elements, the strong relativistic spin-orbit coupling (SOC) effect is an important consideration in describing the physical properties of 5$d$ TMOs, and it rigorously competes with other relevant energies, particularly the on-site Coulomb repulsion $U$ and crystal electric field (CEF) interactions arising from the surrounding coordinate environment [3]. Consequently, a variety of exotic quantum phases, including the $J_{eff}$ =1/2 Mott state [1, 2], superconductivity [4, 5], a topological insulator [6, 7], Weyl semimetal with Fermi arcs [8], are shown and expected in 5$d$ TMOs.

In the case of the most well-known 5$d$ TMOs, the layered $Sr_2IrO_4$ known as SOC-related Mott insulator, has been subjected to the most extensive investigations due to its structural and electronic similarities to those of the celebrated $La_2CuO_4$ [9]. $Sr_2IrO_4$, which has the $K_2NiF_4$ structure, is based on $Ir^{4+}$ in $IrO_6$ octahedra [10]. The octahedral CEF splits the 5$d$ bands into $t_{2g}$ and $e_g$ states, and the $t_{2g}$ manifold splits into two effective angular momentum energy levels (a doublet $J_{eff}$ = 1/2 and a quartet $J_{eff}$ = 3/2) *via* the strong SOC [1, 2]. Since the $Ir^{4+}$ ions provide five 5$d$-electrons, four of them fill the lower $J_{eff}$ = 3/2 bands, and one electron partially fills the $J_{eff}$ = 1/2 band where the Fermi level $E_F$ resides. The $J_{eff}$ = 1/2 band is so narrow that even a reduced $U$ is sufficient to open a gap, supporting an insulating behavior [1, 2]. Accordingly, the delicate competition among the various interactions such as the CEF,

SOC and $U$ plays a crucial role in realizing the $J_{eff}$ = 1/2 intriguing physics for $Sr_2IrO_4$.

Several theoretical investigations predicted that the unconventional high-temperature superconductivity can be induced by carrier doping in this $Sr_2IrO_4$-based system [4, 11-13]. Very recently, experiments on doped $Sr_2IrO_4$ have uncovered tantalizing evidence of a Fermi surface split up into disconnected segments ("Fermi arcs") [14] and a low-temperature gap with d-wave symmetry [15, 16], which are hallmarks of the doped cuprates [17]. However, the true sense of superconductivity has not been found yet, regardless of the electron and hole-doping.

As is known to all, material properties are in a close relationship with their atomic structure. Several reports pointed out that carrier doping can alter lattice parameters and further change the electronic and magnetic behaviors of $Sr_2IrO_4$ [18-21]. Thus, one of the important issues to be resolved for $Sr_2IrO_4$-based system is how carrier doping affect their physical properties in terms of local atomic structure. For instance, Bhatti *et al.* established a relationship between temperature dependent magnetic property and the structural parameters by means of x-ray diffraction (XRD) measurement on $Sr_2IrO_4$ [22]. Furthermore, our pervious investigations suggest the synergistic effect of regular $IrO_6$ octahedra and electron doping that accounts for the transition from a Mott insulator to a conductive state in $Sr_{2-x}La_xIrO_4$-based system [23]. A more accurate and systematic investigation of local atomic displacements induced by carrier doping is still necessary to better understand the $Sr_2IrO_4$ system, that may be beneficial to uncover the potential superconductivity in SOC-related Mott insulators.

Synchrotron-radiation-based X-ray absorption spectroscopy (XAS) is a well-recognized local experimental technique capable of giving substantial information (*i.e.* local atomic and electronic structure) for a material [24, 25]. Moreover, Ir $L_{2,3}$-edge XAS can be competent for providing other valuable information (*e.g.* the strength of SOC) in Ir-based systems [26]. Here, we take the polycrystalline compounds of $Sr_{2-x}IrO_4$ as the object of our research. As the Sr-vacancies approach $x \sim 0.45$, the properties of the material change suddenly in three aspects [27]: 1) an insulator-metal transition; 2) a sudden drop of the localization temperature and the antiferromagnetic transition temperature, together with an anomalous sign reverse of the Curie-Weiss temperature; 3) a significant structural change probed by both XRD and Raman spectra. All these aspects strongly suggest the existence of a critical point around $x \sim 0.45$ in this system. Therefore, an investigation of the critical point is highly desired and might be a key to achieve the potential superconductivity in $Sr_2IrO_4$-based system. In this contribution, detailed investigations of the Sr-vacancy effect at the critical point are probed by a comprehensive analysis of Ir $L_{2,3}$-edge XAS.

## 2. Experiments and Calculations

Polycrystalline compounds with nominal compositions of $Sr_{2-x}IrO_4$ were synthesized by solid-state reaction method as mentioned elsewhere [27]. The samples were well characterized for their phase purity prior to the XAS measurements. We collected both the Ir $L_2$-edge x-ray absorption near edge structure (XANES) and $L_3$-edge extended x-ray absorption fine structure (EXAFS) spectra of $Sr_{2-x}IrO_4$ at room temperature, which occur at energy of 12.824keV and 11.215keV, respectively. The XAS spectra were collected at the BL-14W1 beamline of Shanghai Synchrotron Radiation Facility (SSRF). Several scans were collected to ensure the spectral reproducibility. The storage ring was working at electron energy of 3.5 GeV, and the maximum stored current was about 250mA. The energy of the incident energy was tuned by scanning a Si (111) double crystal monochromator with energy resolution about $10^{-4}$. Data reduction was performed using the IFEFFIT program package [28].

Density functional theory (DFT) calculations for $Sr_{2-x}IrO_4$ system were carried out within the local-density approximation (LDA) including SOC and on-site Coulomb $U$ using OPENMX [29], based on the linear-combination-of-pseudo-atomic-orbitals (LCPAO) method [30]. SOC is treated *via* a fully relativistic j-dependent pseudopotential in the non-collinear DFT scheme, with 300 Ry of energy cutoff and 11×11×11 k-grid. For $Sr_{2-x}IrO_4$, only one and two of the eight Sr atoms were removed in a unit cell, which correspond to $x = 0.25$ and 0.5, respectively. Atomic positions of the structure have been relaxed until all the force components became smaller than $10^{-4}$ eV/a.u. The on-site Coulomb interaction parameter $U$ is set to 2.0 eV for Ir $d$-orbitals in our LDA + SOC + $U$ calculations.

## 3. Results and Discussions

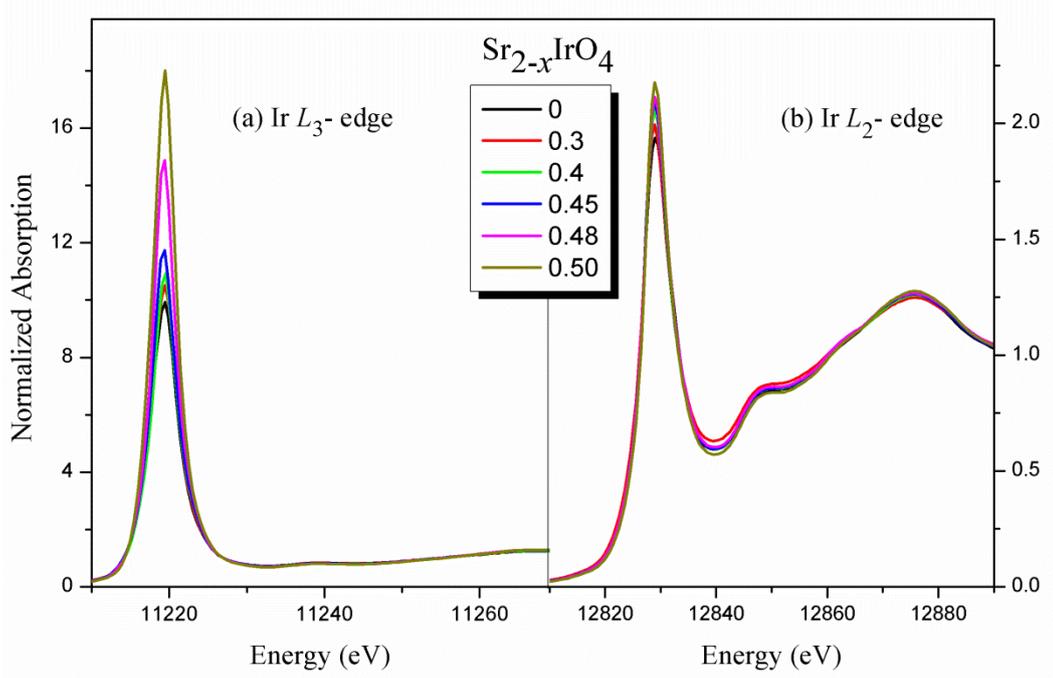

**Fig. 1.** Ir $L_{2,3}$-edge normalized XANES spectra of $Sr_{2-x}IrO_4$.

The Ir $L_{2,3}$-edge XANES spectra of $Sr_{2-x}IrO_4$ compounds are provided in Fig. 1. The most striking characteristic is the prominent white-line (WL) features which correspond to $2p \rightarrow 5d$ electronic transitions. From the dipole selection rules, $\Delta J$ must be equal to 0 or $\pm 1$, it follows that for $Sr_2IrO_4$-based system the Ir $L_2$-edge will be sensitive to transitions involving $5d_{3/2}$ (*i.e.* $J_{eff} = 3/2$) holes, while the $L_3$-edge will be related to both $5d_{5/2}$ (upper Hubbard band associated with $J_{eff} = 1/2$ and the CEF $e_g$ manifolds) and $5d_{3/2}$ final states. Because the WL features are more pronounced at the $L_3$ edge than the $L_2$ edge, we can also infer that these unoccupied $5d$ states are primarily $5d_{5/2}$ rather than $5d_{3/2}$ in nature, consistent with the almost completely filled $J_{eff} = 3/2$ bands in $Sr_2IrO_4$ [2]. Furthermore, the integrated intensity of the WL features is proportional to the local density of unoccupied final states (*i.e.* the population of $5d$ holes) in the system. From Fig. 1 the integrated intensity of WL features at both absorption edges grows steadily with increasing the Sr-vacancies, pointing out the increment of Ir-5$d$ holes in Sr$_{2-x}$IrO$_4$ compounds. To demonstrate the above discussion, the DFT calculations for Sr$_{2-x}$IrO$_4$ ($x$ = 0, 0.25 and 0.5) were performed. In Table 1, we summarized the number of electrons for each element. It is found that the holes induced by Sr-vacancies are located around all atoms, especially the Ir atoms, in agreement with the result of XANES spectra.

**Table 1**

The number of electrons belonging to each atomic species for Sr$_{2-x}$IrO$_4$.

| Sr$_{2-x}$IrO$_4$ | Ir | Sr | O |
|---|---|---|---|
| $x = 0$ | 15.91 | 8.46 | 6.54 |
| $x = 0.25$ | 15.84 | 8.42 | 6.51 |
| $x = 0.5$ | 15.52 | 8.38 | 6.48 |

The relative strength of the WL features at the $L_3$ and $L_2$ edges (also known as the branching ratio BR = $I_{L_3}/I_{L_2}$) is an important quantity [26], since it is directly related to the expectation value of the angular part of the SOC of 5d states *via* BR = (2+$r$)/(1-$r$), where $r$ = ⟨L·S⟩/$n_h$ and $n_h$ is the average number of 5$d$ holes [31, 32]. For a system with negligible SOC the statistical branching ratio from XANES is 2. To discuss the branching ratio on a quantitative level, the integrated intensity of WL feature at each edge was provided based on the least-squares method as mentioned elsewhere, using a Lorentzian + arctangent fit function [23]. By analysing the XANES spectra, the branching ratio for Sr$_2$IrO$_4$ is BR = 6.45, a value which is several times larger than the statistical one, indicating the presence of extremely large SOC effects. Using $n_h = 5$ we obtain $\langle L \cdot S \rangle = 2.99$ in units of $\hbar^2$, consistent with other reports [26].

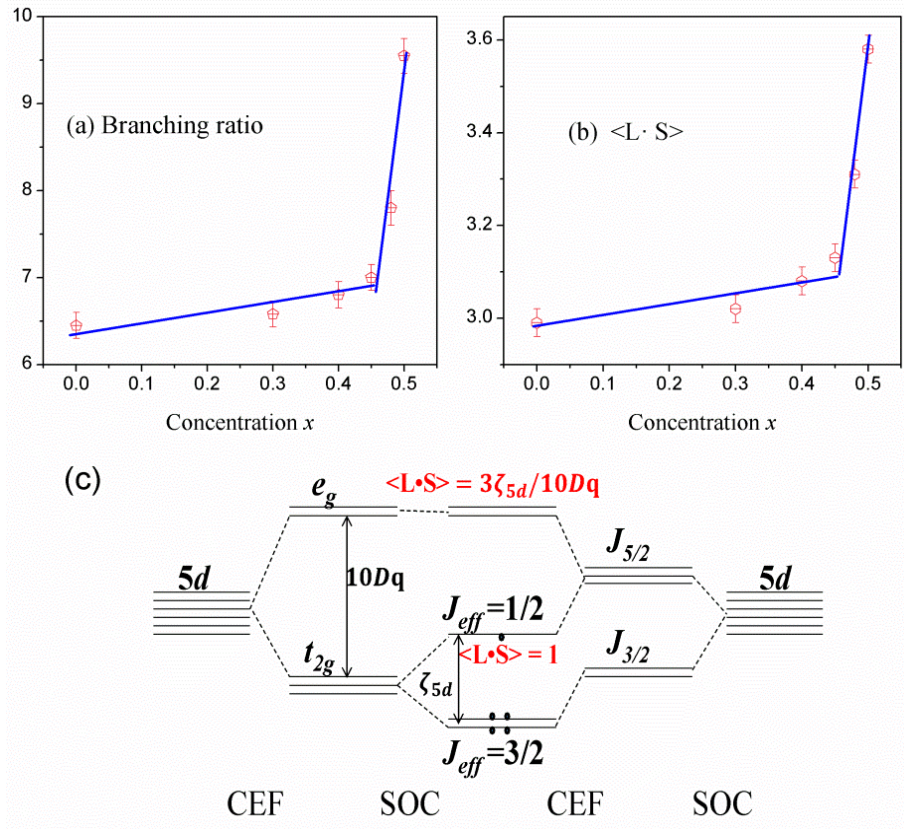

**Fig. 2.** The branching ratio $I_{L_3}/I_{L_2}$ (a) and $\langle L \cdot S \rangle$ (b) as a function of Sr-vacancy concentrations. (c) A schematic energy diagram of 5$d$ level splittings by the CEF and SOC, based on Ref. 2.

As shown in Fig. 2, both the branching ratio and $\langle L \cdot S \rangle$ increase gradually with the increasing Sr-vacancies, while a sudden rise occurs in the deep doping region around $x = 0.45$. This anomalous kink reminds us about the critical point in $Sr_{2-x}IrO_4$ system, implying potential changes of the fundamental local atomic structure at the critical point. The $\langle L \cdot S \rangle$ is a property of the local moment, it is mostly determined by the

SOC and the CEF acting on 5$d$ electrons. Note that since XANES probes all empty 5$d$ states, the measured ⟨L•S⟩ will be the sum of two contributions (illustrated in Fig. 2(c)): a single hole in the $J_{eff}$ = 1/2 state (⟨L•S⟩ = 1) and four holes in the $e_g$-derived states (⟨L•S⟩ = 4 × 3$\zeta_{5d}$/10$D$q) [33, 34], where $\zeta_{5d}$ represents the SOC of the 5$d$ states and 10$D$q is the parameter of octahedral CEF. Therefore, the Sr-vacancies can effectively alter the relative strength of the SOC and CEF, which could be useful to describe the fundamental properties of $Sr_{2-x}IrO_4$ system.

It is well known that Ir atoms play a significant role in extremely strong SOC effects of $Sr_2IrO_4$-based system, while the cation substitutions can change the strength of SOC [35]. For instance, a substitution of Ir with Ru/Rh would tune the SOC strength from the strong 5$d$ regime to the moderate 4$d$ regime [35]. On the other hand, the substitution at Sr sites (*e.g.* La-doping) just tunes the band filling, and does not alter the SOC strength [35]. That is to say, only the substitution at Ir sites can significantly tune the SOC strength of the system. Roughly speaking, in our study Sr-vacancies impose negligible effect on the SOC strength, and we neglect the SOC changes in our discussion. Thus we will focus on the CEF effect *versus* Sr-vacancies.

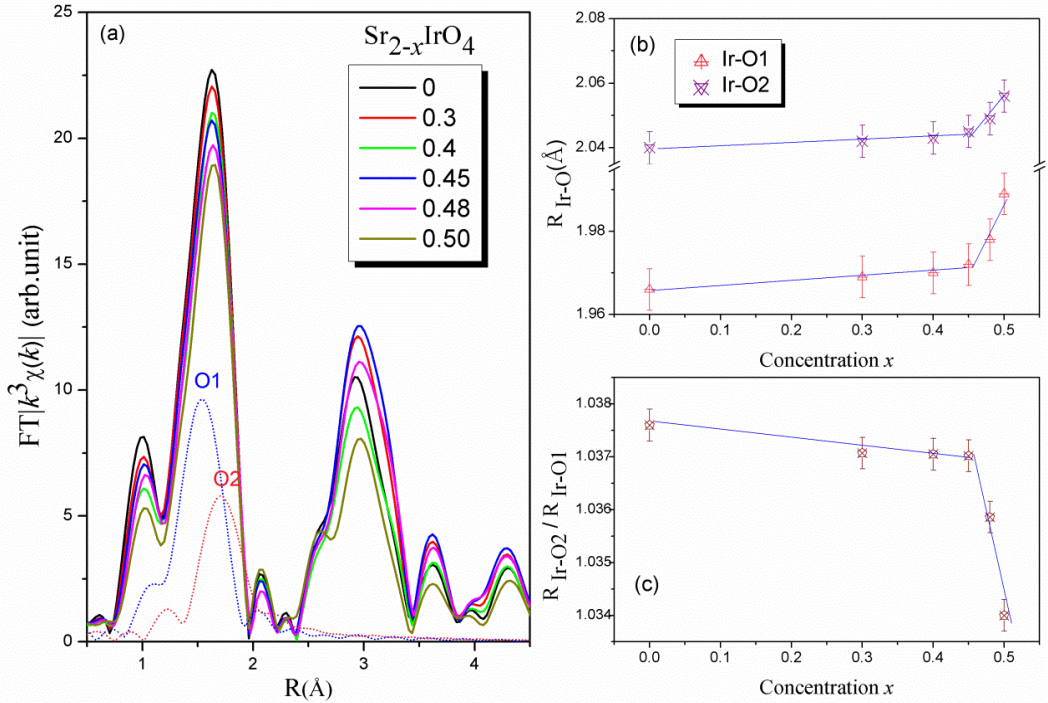

**Fig. 3.** (a) Fourier transform magnitudes of $k^3\chi(k)$ EXAFS signals of $Sr_{2-x}IrO_4$ compounds at Ir $L_3$-edge and the contributions of Ir-O1/O2 pairs calculated by FEFF code. The Results from the fit of Ir-O1/Ir-O2 bond distances are shown in panel (b). (c) The derived differences between the distances of Ir-O1 and Ir-O2 bonds as a function of Sr-vacancy concentrations. Error bars correlated to the uncertainty value are given.

In the $Sr_2IrO_4$-based system, the CEF interaction is arisen from surrounding six oxygen atoms in a nearly octahedral local coordinate environment [10]. The $IrO_6$ octahedron contains two distinct oxygen positions (four basal O1 and two apical O2), which is slightly elongated apically. Therefore, to gain an insight into the local atomic displacements induced by Sr-vacancies, we have undertaken detailed structural study (*e.g.* the bond distances of Ir-O1 and Ir-O2) by means of the Ir $L_3$-edge EXAFS spectra. The Fourier transform (FT) magnitudes of the EXAFS oscillations providing real space information around Ir atoms are plotted in Fig. 3(a). Note that the positions of the peaks in the FT are shifted a few tenths of Å from the actual interatomic distances because of the EXAFS phase shift [36]. Obviously, large changes in the FTs of Ir $L_3$-edge can be seen, indicating the atomic displacements induced by Sr-vacancies.

The EXAFS amplitude depends on several factors and is given by the following general equation [37]:

$$\chi(k) = \sum_j \frac{N_j S_0^2}{k R_j^2} f_j(k, R_j) \exp[-2k^2 \sigma_j^2] \exp[\frac{-2R_j}{\lambda}] \sin[2kR_j + \delta_j(k)] \tag{1}$$

where $N_j$ is the number of neighboring atoms at a distance $R_j$, $S_0^2$ is the passive electron reduction factor, $f_j(k, R_j)$ is the backscattering amplitude, $\lambda$ is the photoelectron mean free path, $\delta_j(k)$ is the phase shift and $\sigma_j^2$ is the correlated Debye-Waller factor.

In order to obtain quantitative results, we fit the first peak of EXAFS spectra ($R$ = 1.0 ~ 2.2 Å) involving contributions of four basal O1 and two apical O2, which were isolated from the FTs with a rectangular window. The range in $k$ space was 3 ~ 15 Å$^{-1}$ and that in $R$ space was 1.0 ~ 2.2 Å. For the least-squares fits, average structure measured by diffraction on $Sr_2IrO_4$ system is used as the starting model [27]. The backscattering amplitudes and phase shift were calculated using the FEFF code [38]. Only the radial distances $R_j$ and the corresponding $\sigma_j^2$ were allowed to vary, with coordination numbers $N_j$ fixed to the nominal values. The passive electrons reduction factor $S_0^2$ and photoelectron energy zero $E_0$ were also fixed after fit trials on different scans. The number of independent parameters which could be determined by EXAFS is limited by the number of the independent data points $N_{ind}$ ~ $(2\Delta k \Delta R)/\pi$, where $\Delta k$ and $\Delta R$ are respectively the ranges of the fit in the $k$ and $R$ space [37]. In our case, $N_{ind}$ is 9 ($\Delta k = 12$ Å$^{-1}$, $\Delta R = 1.2$ Å), sufficient to obtain all parameters.

As shown in Fig. 3(b), the distances of both Ir-O1 and Ir-O2 pairs increase gradually with the increasing Sr-vacancies, while an obvious kink occurs in the deep doping region. Note that the IrO$_6$ octahedron is slightly elongated apically, and deep Sr-vacancies suddenly decrease the ratio of apical Ir-O2 to basal Ir-O1 distance (illustrated in Fig. 3(c)), indicating less distorted IrO$_6$ octahedra induced by deep Sr-vacancies. Surprisingly, all these structural changes are in coincidence with the existence of the critical point around $x = 0.45$. That is to say, the changes of fundamental atomic structure would be important to describe the properties around the critical point.

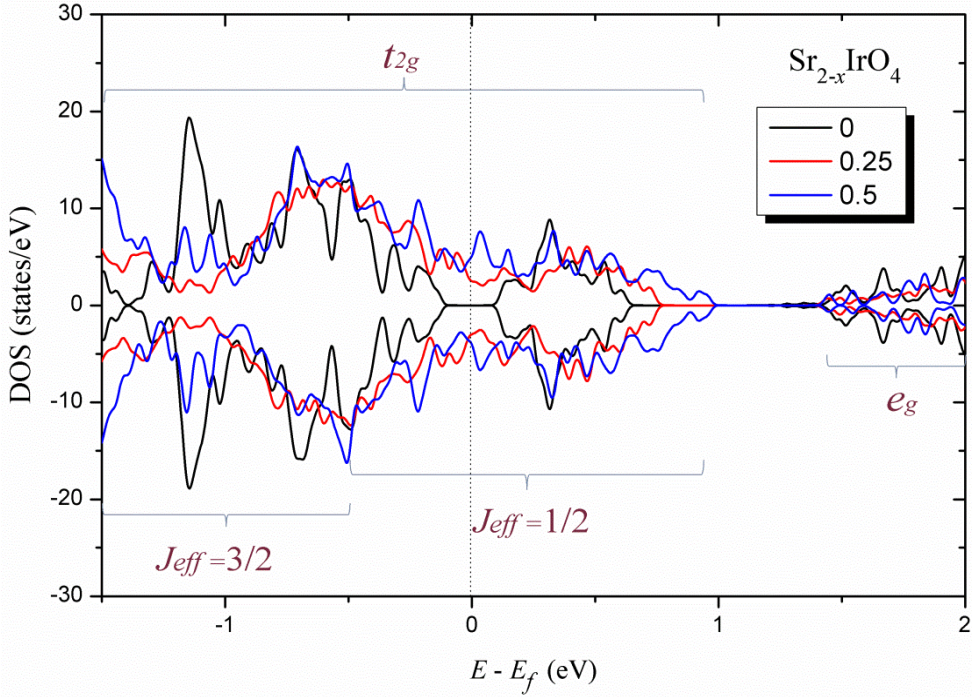

**Fig. 4.** Total density of states for Sr$_{2-x}$IrO$_4$ system.

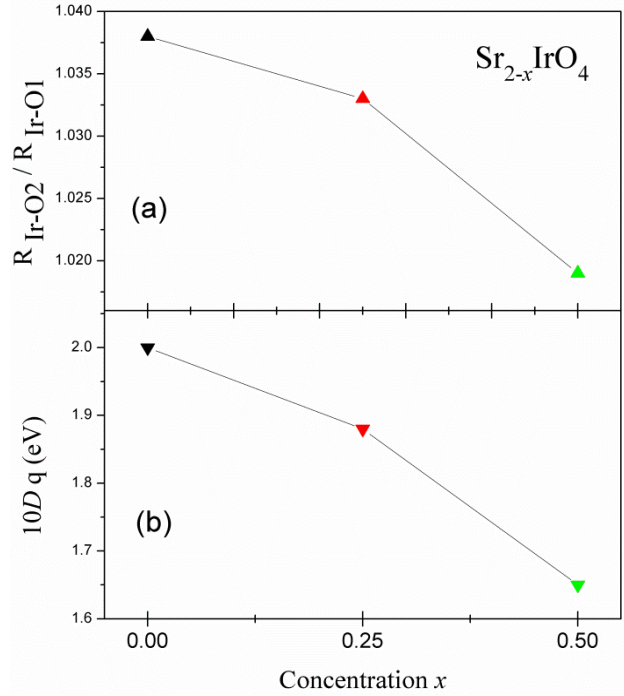

**Fig. 5.** The ratio of apical Ir-O2 to basal Ir-O1 bond length (a) and the CEF parameter $10Dq$ (b) for $Sr_{2-x}IrO_4$ system.

In order to validate the above XAS results, we performed DFT calculations on $Sr_{2-x}IrO_4$ system. The total density of states in Fig. 4 suggests that deep Sr-vacancies turn the $Sr_{2-x}IrO_4$ system from a Mott insulator to the metallic state, in agreement with the experimental results [27]. As shown in Fig. 5, Sr-vacancies also induce structural changes driving the ratio of apical Ir-O2 to basal Ir-O1 bond length towards the value of 1, suggesting the less distorted $IrO_6$ octahedra. Accordingly, the $10Dq$ parameter decreases suddenly (*i.e.* the weakened CEF) along with the deep Sr-vacancies, consistent with the abrupt enhancement of ⟨L•S⟩.

As mentioned above, our results provide experimental and calculated data on the less distorted $IrO_6$ octahedra induced by deep Sr-vacancies; this in turn tunes the relatively strength of the SOC and CEF for $Sr_{2-x}IrO_4$ system, which would dictate the abnormal behavior of electronic and magnetic properties at the critical point.

## 4. Conclusion

To summarize, we have investigated the Sr-vacancy effect on the spin-orbit coupled $Sr_2IrO_4$ probed by Ir $L_{2,3}$-edge XAS. By fitting the white line features of Ir $L_{2,3}$-edge XANES, the sudden enhancement of both the branching ratio and the derived <L•S> were observed in the vicinity of the critical point. Further investigations in terms of local structure showed that the abrupt rise of <L•S> near the critical point is intimately related to the less distorted $IrO_6$ octahedra (*i.e.* the weakened CEF). So in this contribution we establish the delicate interplay between the critical point and the regular configuration of $IrO_6$ octahedra, hoping this point of view could be beneficial to our understanding of abnormal physics properties in $Sr_{2-x}IrO_4$-based system.

## Acknowledgement

This work was partly supported by the National Natural Science Foundation of China (NSFC 11405089 and 11504182), the "Six Talents Peak" Foundation of Jiangsu Province (2014-XCL-015), the Nanotechnology Foundation of Suzhou Bureau of Science and Technology (ZXG201444).